\documentclass[9pt]{article}
\usepackage{spconf}
\usepackage{amssymb}
\usepackage[pdftex]{graphicx}
\usepackage[cmex10]{amsmath}
\usepackage{subfigure}
\usepackage{times}
\usepackage{comment}
\usepackage{multirow}
\usepackage{ifpdf}
\usepackage{blindtext,subfig}
\usepackage{amsfonts}
\usepackage{txfonts}
\usepackage{cite}
\usepackage{listings}
\usepackage{xcolor}
\usepackage[]{algorithm2e}
\usepackage{url}
\makeatletter
\def\url@leostyle{%
  \@ifundefined{selectfont}{\def\UrlFont{\sf}}{\def\UrlFont{\small\ttfamily}}}
\makeatother

\DeclareMathOperator*{\argmin}{\arg\!\min}

\title{A two-stage video coding framework with both self-adaptive redundant dictionary and adaptively orthonormalized DCT basis}
\name{Yuanyi Xue, Yi Zhou, and Yao Wang}
\address{Department of Electrical and Computer Engineering\\
        Polytechnic School of Engineering, New York University \\
        Email: yxue@nyu.edu,~yi.zhou@nyu.edu,~yw523@nyu.edu}

\begin{document}
\ninept
\maketitle
\begin{abstract}
In this work, we propose a two-stage video coding framework, as an extension of our previous one-stage framework in~\cite{xue2014video}. The two-stage frameworks consists two different dictionaries. Specifically, the first stage directly finds the sparse representation of a block with a self-adaptive dictionary consisting of all possible inter-prediction candidates by solving an L0-norm minimization problem using an improved orthogonal matching pursuit with embedded orthonormalization (eOMP) algorithm, and the second stage codes the residual using DCT dictionary adaptively orthonormalized to the subspace spanned by the first stage atoms. The transition of the first stage and the second stage is determined based on both stages' quantization stepsizes and a threshold. We further propose a complete context adaptive entropy coder to efficiently code the locations and the coefficients of chosen first stage atoms. Simulation results show that the proposed coder significantly improves the RD performance over our previous one-stage coder. More importantly, the two-stage coder, using a fixed block size and inter-prediction only, outperforms the H.264 coder (x264) and is competitive with the HEVC reference coder (HM) over a large rate range.
\end{abstract}

\section{Introduction}\label{sec:intro}
The current state-of-art coding method like HEVC~\cite{HEVCreview} usually relies on a block based hybrid scheme. A block in a video frame is first predicted through the best inter- or intra-predcition, and the prediction error (residual) is then passed into a transform coder, where a fixed orthogonal transform such as DCT is used. The resulting transform coefficients are then quantized and the non-zero DCT coefficients, as well as the prediction candidate locations (e.g. the motion vectors in the case of inter-prediction) and/or modes are finally coded using an entropy coder. In~\cite{xue2014video}, we proposed a new video coding scheme that directly represents each block in a video frame using a self-adaptive dictionary consisting of all possible temporal prediction candidate blocks over a search range in the previous frame. Here in this work, we propose an extension of that by including a second stage that codes the residual after approximating the current block using the chosen atoms in the first stage. The rationale for this second stage is that we have identified the self-adaptive dictionary loses its representation efficiency after first few chosen atoms. This is because the atoms in the self-dictionary may be highly correlated and concentrated in a sub-space. We code the residual after the second stage using a non-redundant orthonormal dictionary that spans the null-space of the sub-space spanned by chosen atoms in the first stage. In our current implementation, we derive this second dictionary by orthonormalizing all DCT transform basis blocks with respect to the chosen atoms in the first stage.

We would like to emphasize that the proposed two-stage framework differs significantly from the state-of-art video coding standards. Specifically, although our self-adaptive dictionary consists of conventional motion search candidate blocks and we use a linear combination of a few motion search candidates for the first stage representation (which we term as atoms as in sparse recovery literature), we allow any combination of possible candidates and the weights are solved through an optimization problem; we do not restrict the number of chosen atoms to a pre-set number, but minimizing this number through minimizing the L0-norm of the weights. By comparison, state-of-art coding standards like HEVC only permit a fixed linear combination of several adjacent motion candidates through the use of fractional-pel motion compensation, where the weights are implicitly determined by the interpolation filter for deriving fractional-pel motion vectors. Besides, our second stage orthonormalizes the fixed DCT basis with respect to the block-dependent first stage chosen atoms. Effectively, the orthonormalization in the second stage adaptively alters the DCT basis to span over the null space of the first stage reconstruction and hence is more efficient in representing the residuals. To the contrary, the transform coding stage in all current video coding standards is fixed.

Several research groups have attempted using redundant dictionaries for block-based image and video coding, including~\cite{KSVD_ImgComp,Guillemot_JSTSP_2011,Zakhor_Dictionary4vid_CSVT2004,MERL_ISCAS2011,SparseDCT_TIP2013}.  In all reported dictionary-based video coders,  the dictionary atoms are used to represent the motion-compensated residual block for intra/inter-frame video coding. Therefore, they are very different from what is proposed here. The work in~\cite{TsinghuaOnlineDictionary} codes each frame in the intra-mode, with a dictionary that is updated in real time based on the previously coded frames.  Although such online adaptation can yield a dictionary that matches with the video content very well, it is computationally very demanding. Furthermore, the dictionary is not block adaptive. The proposed framework uses a self-adaptive dictionary that only depends on the block location, without requiring realtime design/redesign of the dictionary. 

Two major challenges have to be addressed for making the two-stage framework efficient. The first is the fact that the redundant dictionary is highly correlated. Directly quantizing the coefficients with the highly correlated atoms is inefficient. In this work, as our original work in~\cite{xue2014video}, we recursively orthonormalizes the chosen atoms during the iteration of OMP. However we have developed an improved OMP algorithm to solve the sparse recovery problem, which can yields a sparser representation than the original OMP method~\cite{xue2015revomp}. The coefficients associated with the orthonormalized chosen atoms can be quantized independently without losing efficiency. The second challenge is how to determine the switch point from the first stage to the second stage. In this work, we switch to the second stage when the first stage self-adaptive dictionary begins to slowly correct the residual, which indicates that the remaining atoms in the self-adaptive dictionary have limited representation power for the remaining residual.

In the remainder of this paper, we describe the proposed two-stage framework in Sec.~\ref{sec:spdcp}; we present the entropy coder design in Sec.~\ref{sec:ent}; in Sec.~\ref{sec:res} we show the RD performance and comparisons. We conclude the paper in Sec.~\ref{sec:con}.
\section{Two-stage Video Coding Framework}\label{sec:spdcp}
\begin{figure}
\centering
\includegraphics[width=1.05\linewidth]{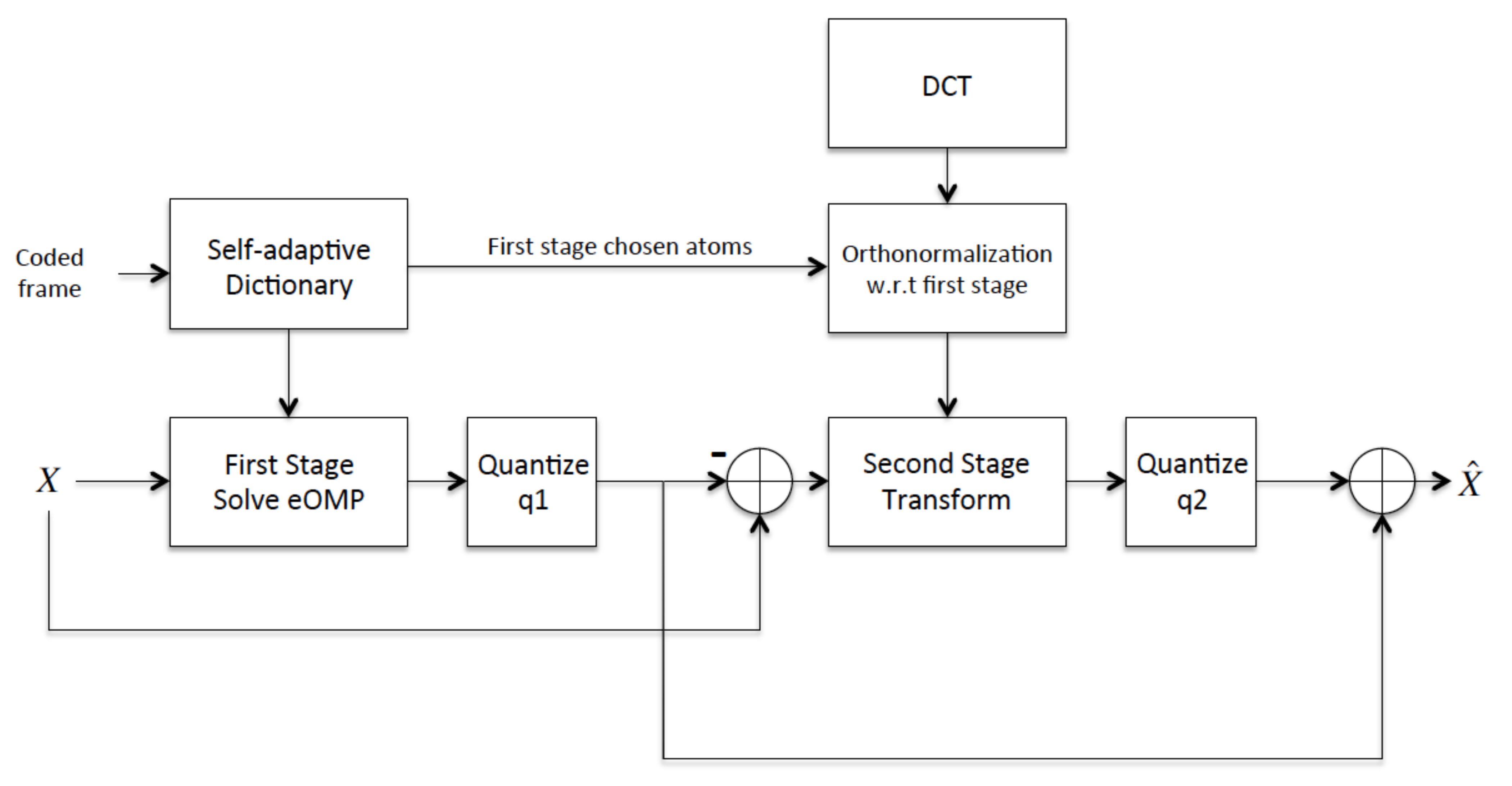}
\caption{Proposed two-stage framework}
\label{fig:Framework}
\end{figure}

The proposed two-stage framework is illustrated in Fig.~\ref{fig:Framework}. We describe the different components in more details below.
\subsection{First stage using self-adaptive dictionary with eOMP}
The first stage is similar to the one reported in~\cite{xue2014video}. For each mean-removed block represented as a vector $\mathrm{x} \in \mathbb{R}^{N}$, denote the dictionary by $\mathbf{D} \in \mathbb{R}^{N\times M}$, which consists of the mean-removed vector representation of all possible motion search candidate blocks corresponding to integer motion vectors in a preset motion search range, and the coefficients by $\mathrm{w} \in \mathbb{R}^{M}$. We find the sparse set of coefficients by solving the following L0-norm minimization problem.
\begin{align}\label{eq:l0}
\argmin_{\mathrm{w}}\quad &|\mathrm{w}|_0 \nonumber \\
\text{s.t.} \quad \|\mathrm{x} - \mathbf{D}\mathrm{w}\|_2^2 \leq N\epsilon^2 & \cup \text{Early termination constraint.}
\end{align}
We set $\epsilon^2$ to be slightly larger than the expected distortion introduced by the quantization stepsize $q_1$, i.e. $\epsilon^2 > q_1^2/12$. 

Because of the non-convexity of L0-norm, greedy algorithm is often used to solve this problem. One prominent algorithm is the orthogonal matching pursuit (OMP). In the original OMP algorithm~\cite{OMP07}, at every iteration, the correlation between the residual and the remaining atoms are evaluated, and the best correlated atom is chosen; a least square step is then used to update all the coefficients, where the coefficients are related to the original chosen atoms. The least squares update step effectively makes the residual to lie in the null space of the chosen atoms. In~\cite{xue2014video}, we implemented the original OMP algorithm with embedded orthonormalization of chosen atoms so that we could solve the sparse reconstruction and atom orthonromalization simultaneously. This is accomplished by orthonormalizing the latest chosen atom with respect to all previously chosen atoms using a Gram-Schmidt procedure, and finding the coefficient associated with the orthonormalized latest chosen atom by a simple inner product. In this work, we use an improved version of OMP, denoted by eOMP~\cite{xue2015revomp}. In each iteration, instead of evaluating the correlation between the residual and the remaining atoms, we first update the remaining atoms by orthonormalizing them to all previously chosen orthonormalized atoms. This orthonormalization can be done efficiently through a recursive one-step orthonormalization procedure. Let $\mathrm{a}_i$ denote the $i$-th remaining atoms, $\mathrm{b}$ the last chosen atom, the one-step orthonormalization is given by the following:
\begin{align}\label{eq:1steporth}
& \mathrm{a}_i \gets \mathrm{a}_i - \langle \mathrm{b},\mathrm{a}_i \rangle \mathrm{b} \nonumber \\
& \mathrm{a}_i \gets \frac{\mathrm{a}_i}{\lVert\mathrm{a}_i\rVert_2}
\end{align}
The correlation between the residual and orthonormalized remaining atoms are then evaluated and the atom with the largest correlation (i.e. the magnitude of the inner product) is chosen. Note that this chosen atom is already orthonormalized and the coefficient corresponding to this atom has already been calculated. By recursively updating the remaining atoms, we have shown in~\cite{xue2015revomp} that this improved eOMP has better sparsity recovery capability than the original OMP algorithm.

The early termination constraint in Eq.~\ref{eq:l0} is expressed in the terms of residual norm reduction ratio threshold $t$, a threshold on the ratio between the residual norm difference of the current and last iterations and the residual norm of the last iteration. This constraint is specially designed to terminate the OMP algorithm before reaching the fidelity constraint if newly chosen atom does not decrease the residual norm significantly, which usually signifies the low correlation of the residual and the remaining atoms and the loss of the representation power of the remaining atoms in the self-adaptive dictionary. Obviously, the choice of $t$ will affect the overall coding efficiency and must be chosen appropriately.

Once all the orthonormalized atoms denoted by a matrix $\mathbf{B}$ and their corresponding coefficients $\mathrm{c}$ are found, we apply uniform quantization with deadzone to each coefficient with the same stepsize $q_1$, with the deadzone $\Delta=q_1/6$, following the HEVC standard recommendation for inter-coded block.
%

Denoting the quantized coefficients by $\hat{\mathrm{c}}$, the first stage reconstructed block can be represented by $\hat{\mathrm{x}}_1 = \mathbf{B}\hat{\mathrm{c}}$, and the residual is $\mathrm{r} = \mathrm{x} - \hat{\mathrm{x}}_1$.
\subsection{Second stage using DCT basis}
\begin{figure}
\centering
\includegraphics[width=0.65\linewidth]{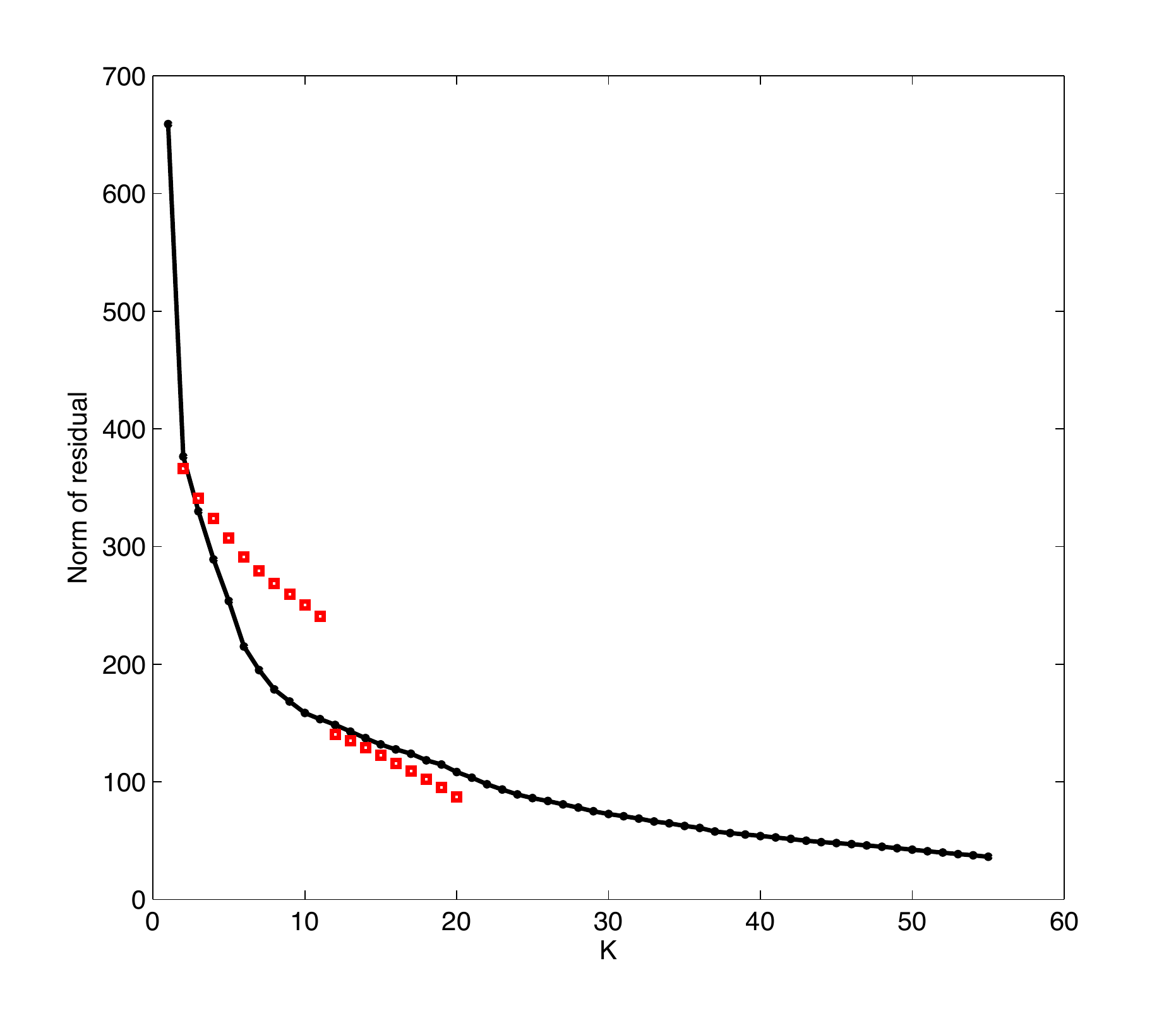}
\caption{Residual norm vs. $K$, comparing using the self-adaptive atoms alone, vs. using two stage.}
\label{fig:2stageprof}
\end{figure}

We start our discussion with a plot demonstrating the motivation for designing the second stage. In Fig.~\ref{fig:2stageprof}, the continuous black curve shows the residual norm versus number of chosen atoms ($K$) from a particular block in sequence \textit{touchdownpass}, when only self-adaptive first stage is used. The beginning part of the curve shows a very fast decay of the residual norm, which indicates the effectiveness of the self-adaptive dictionary. However, after few iterations, the decaying rate significantly slows down for each newly chosen atom. This suggests that the remaining atoms after a few iteration may lie  largely in the same subsapce as the chosen atoms  and cannot represent efficiently the residual, which is orthogonal to this subspace. In other words, the self-adaptive dictionary may not span over the full space $\mathbb{R}^N$ and it is very inefficient to reduce the residual after a few iterations. In addition to the continuous black curve, we also show two red curves originating from two specific $K$s. They represent the residual norm reductions using DCT basis had the first stage stopped at that $K$. In the left red curve, it is apparent that stopping first stage too early yields suboptimal result, i.e. at that stage, self-adaptive dictionary is still more efficient. Interestingly, the right red curve shows an opposite trend, that using DCT basis becomes more efficient in reducing the residual norm compared to continuing with the self-adaptive dictionary. We would like to note that the horizontal axis of Fig.~\ref{fig:2stageprof} is in terms of $K$ rather then the bit rate used to represent $K$ coefficients. Considering the huge size difference between the self-adaptive dictionary (in our experiments, the number of atoms is 2304) and DCT dictionary (256 for $16\times 16$ block size), greater rate savings will be obtained by switching to DCT at the starting point of the right red curve.

Now we give the description of the second stage coder. Given the residual $\mathrm{r}$, the DCT basis vectors are first orthonormalized with respect to the first stage chosen atoms. The coefficients of the residual with respect to these altered DCT basis vectors are then derived simply by performing inner product of the residual vector with each altered DCT basis vector. These coefficients $\lambda$ are quantized with quantization stepsize $q_2$ using the same deadzone quantizer. The residual $\mathrm{r}$ is therefore reconstructed as $\mathbf{T}\hat{\lambda}$, where $\mathbf{T}$ contains the altered DCT basis vectors and $\hat{\lambda}$ are the quantized second stage coefficients. Finally the reconstructed block $\hat{\mathrm{x}}$ is given by $\hat{\mathrm{x}} = \mathbf{B}\hat{\mathrm{c}} + \mathbf{T}\hat{\lambda}$.
\section{Context Adaptive Entropy Coding For the first stage}\label{sec:ent}
The information we need to code in the first stage consists of mainly three parts: 1) the location of the chosen atoms; 2) the order of the chosen atoms, since decoder needs such information to perform the orthonormalization using the same order as the encoder; and 3) the levels of each quantized coefficients. The second stage codes the residual after the first stage using an approach similar to the residual coding method of HEVC~\cite{HEVCreview,CABAC} using DCT, but with a fixed transform block size equal to the block size.

For the first stage, we have developed a new content adaptive arithmetic entropy coder inspired from the CABAC design in HEVC~\cite{HEVCreview}. We would like to emphasize that one major difference between the proposed first stage coder and the CABAC coder in HEVC is that our coder needs to code the order of chosen atoms to enable the decoder to perform the same orthonormalization. As with CABAC in HEVC, we define two probability update modes, the normal mode when the probability for coding each bin is updated based on the past data; and the bypass mode when such probability is fixed. 
\subsection{Coding Atom Locations}
We indicate which atoms are chosen by a 2D binary significance map. For example, if the search range is $-24\cdots 23$, the map size is $48\times 48$, with the top-left element corresponding to the candidate block that is shifted from the current block position by $(-24,-24)$. An element is assigned ``1'' if the corresponding atom is chosen, and ``0'' otherwise. 

\begin{figure}
\centering
\includegraphics[width=0.8\linewidth]{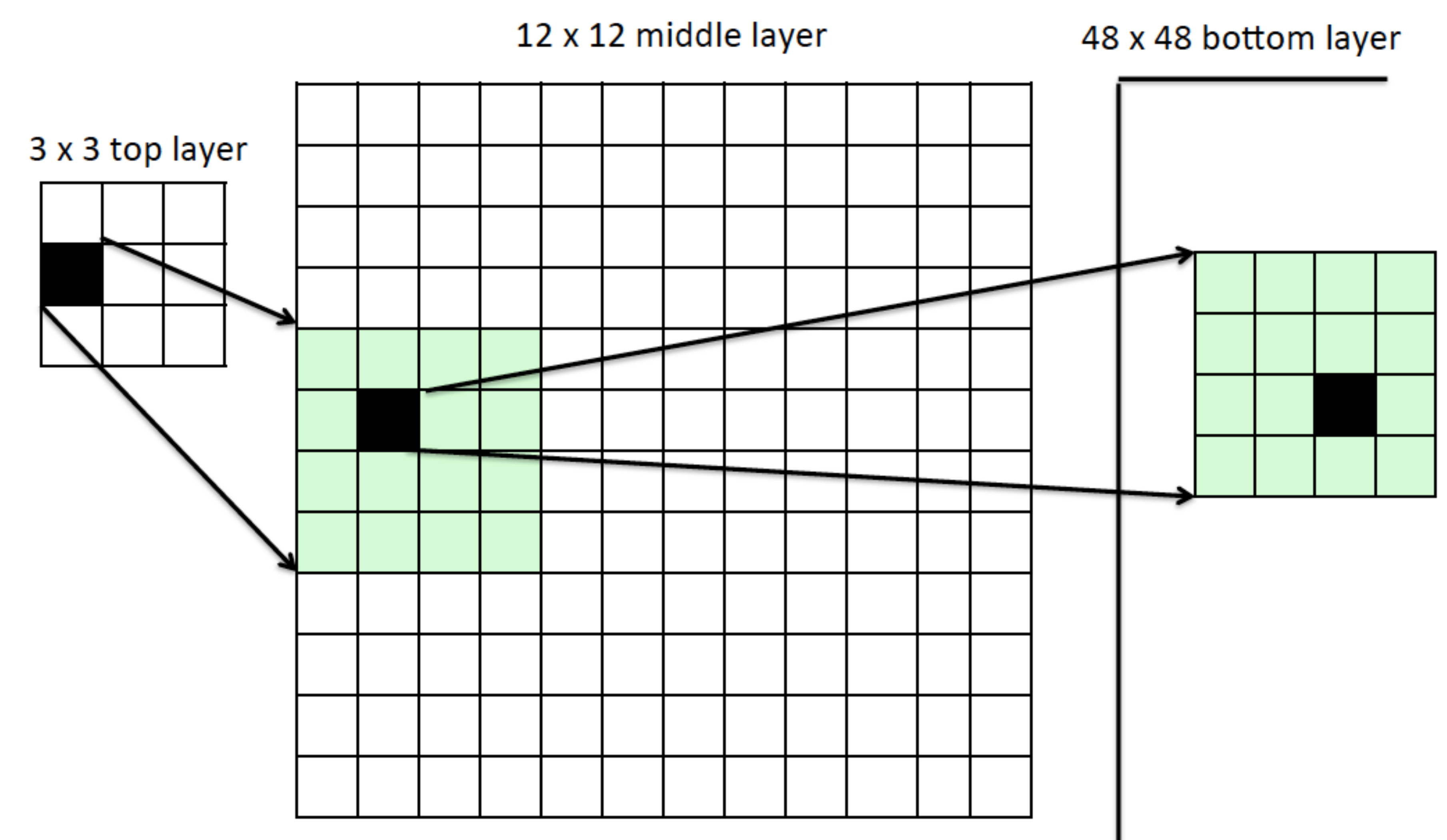}
\caption{Three-layer quad-tree structure for coding significance map.}
\label{fig:3layers}
\end{figure}

We code the significance map using a three-layer multi-resolution approach.  For the search range of $-24 \cdots 23$, the bottom, middle, and top layer have sizes of $48\times 48$, $12\times 12$, and $3\times 3$, respectively, as shown in Fig.~\ref{fig:3layers}. Each element in an upper layer corresponds to a $4\times 4$ region in the next layer. The encoding process starts from the top layer. Every time when a ``1'' is coded, the subsequent $4\times 4$ block should be encoded until the bottom layer is reached.
Every pixel in the top $3\times 3$ layer is coded directly, with a position dependent probability.  In the subsequent layers, for every $4\times 4$ block that corresponds to a ``1'' in the upper layer, the encoder first specifies how many ``1''s are in this block. The coder then specifies the position of the first ``1''  the $4\times 4$ block in a forward horizontal scanning order (i.e. row by row, from left to right) using 4 bits. For the remaining ``1''s in this block, the proposed coder forms a run-length representation, which is binarized using Truncated Unary code and entropy coded using a bitstream using the normal mode. This coding scheme is developed based on the observation that the first stage chosen atom locations are usually sparse and separated. A chosen atom in a particular location often indicates that its nearby atoms are unlikely to be chosen. Therefore, in each $4\times 4$ block, there are usually only one atom chosen.
\subsection{Coding Orders of Chosen Atoms}
We code the orders of the chosen atoms after encoding their locations. Because we can deduce the total number of chosen atoms, i.e. $K$, from the decoded significance map, we adaptively determine the total bits needed to represent the order values. For example, if 3 atoms are chosen, we will form a sequence of 3 values, each represented by 2 bits. After binarization, a bit-plane coding method is employed to code each bit plane of the $K$ order values. The proposed entropy coder encodes the run-lengths of ``1''s in each bit-plane and different bit-planes are encoded separately. Furthermore, given the non-repeatability of the order values, the number of ``1''s is deterministic in each bit-plane. We make use of this property to further reduce the bits spent for coding.
\subsection{Coding Coefficient Levels}
Coefficient levels are encoded in a reversed chosen order, which has been specified in the previous step. We split the levels into the absolute values and signs, and code them separately. For the signs, the bypass mode is always used. For absolute values, predictive coding is used. The last chosen atom's level is predictively encoded first. For remaining levels but the first chosen atom, the coder always uses the previously encoded level as prediction. The first chosen atom's level however, is directly binarized by truncated unary and exponential Golomb code. This is because we have observed a different distribution for the first chosen atoms. Finally all binarized absolute values are coded using the normal mode and written to the bitstream.
\section{Experiments}\label{sec:res}
%
We show the coding performance of the proposed coder and other four comparison coders including the one-stage only baseline coder in~\cite{xue2014video} for two test sequences, \textit{football}, and \textit{city}. The \textit{football} sequence has a frame size $704\times 576$ and the \textit{city} sequence has a frame size $1280\times 720$.

In the proposed two-stage framework, there are a total of three parameters controlling the final rate. Two are the quantization stepsizes for first stage and second stage, $q_1$ and $q_2$, respectively; and the third is the residual norm reduction ratio threshold $t$ for switching from the first stage. At present time, we used an exhaustive search to determine the optimal $q_1$, $q_2$, and $t$ for each target rate in each sequence. We have found that for a given $q_2$, optimal $q_1$ is usually same or slightly smaller than $q_2$, and $t$ increases with $q_1$. 

For both the one-stage and two-stage coders, we use the eOMP proposed in~\cite{xue2015revomp} to derive the first stage atom coefficients. For HEVC, we restrict the prediction and transform unit up to $16\times 16$, but allow the smaller prediction and transform sizes. For H.264, we compare both a version with CABAC and all other advanced coding tools and subpartitions enabled (referred as H264CABAC), as well as the baseline CAVLC with only $16\times 16$ block size, denoted by H264CAVLC.

\begin{figure}
\centering
\subfigure[][]{
\includegraphics[width=0.46\linewidth]{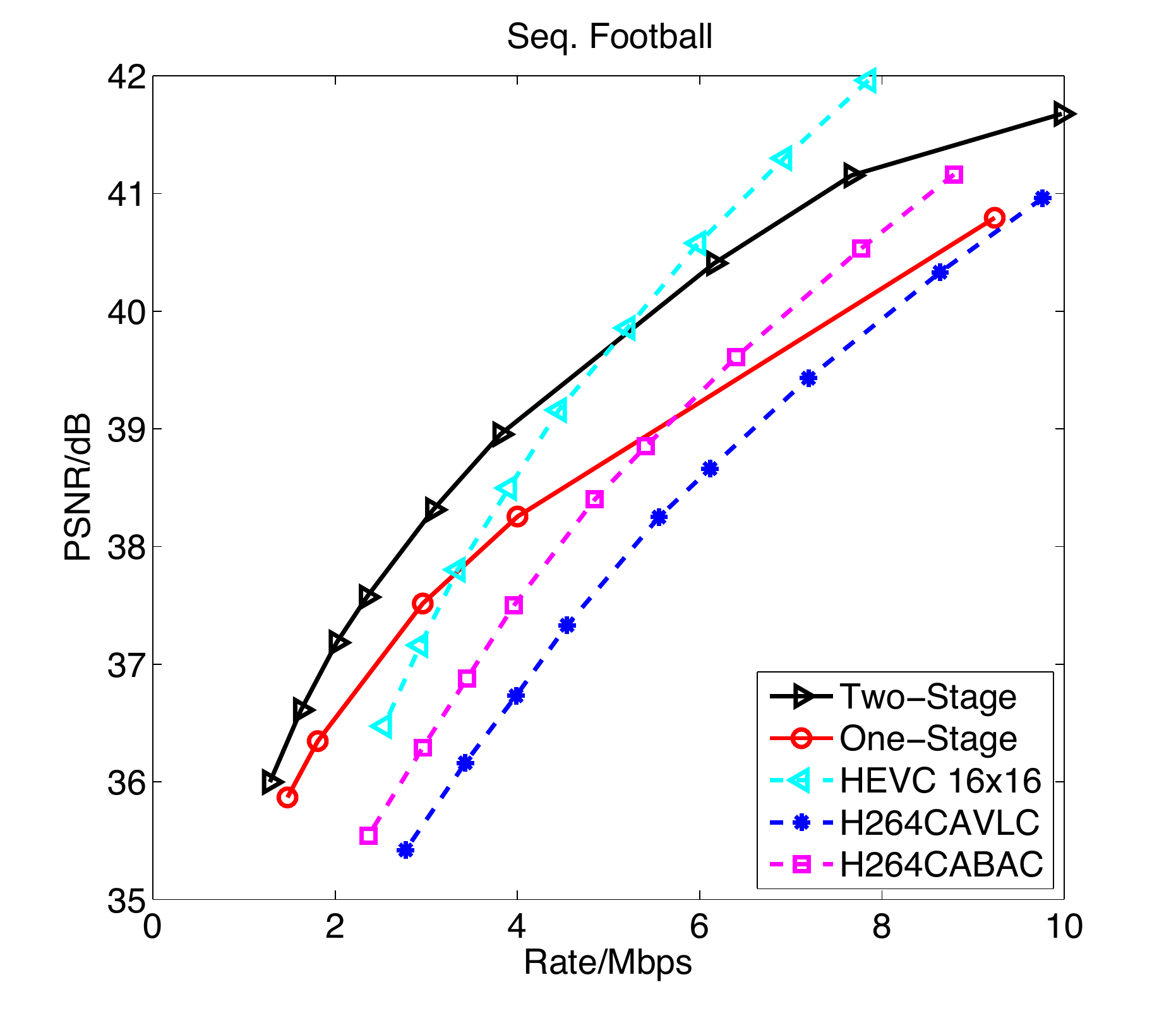}}
\subfigure[][]{
\includegraphics[width=0.46\linewidth]{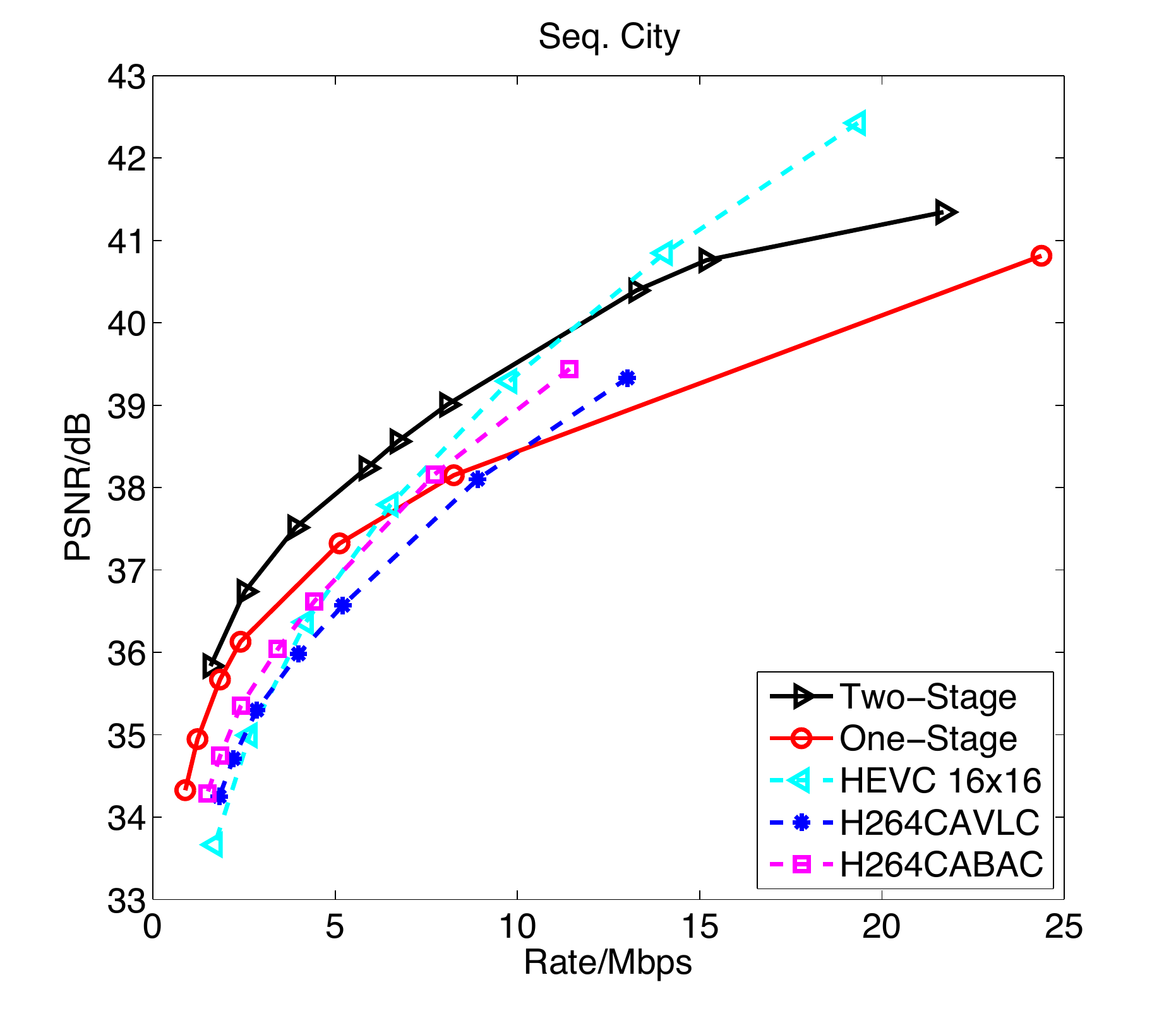}}
\caption{One P-frame rate-distortion curves for proposed and comparison methods. a) Seq. \textit{football}, b) Seq. \textit{city}.}
\label{fig:rd}
\end{figure}

Figure~\ref{fig:rd} shows the rate-distortion curves of encoding one P-frame (based on a previous frame coded as an intra-frame using H.264) for proposed and comparison methods. For both sequences, the two-stage coder significantly improved upon the one-stage coder. It also outperforms both H.264 options significantly. More importantly, it provides substantial gain over HEVC in the low rate range, although it fails to compete with HEVC in the high rate region. We believe one reason our two-stage coder becomes less efficient than HEVC at higher rates is because our second stage coding using a fixed transform size, whereas HEVC allows variable transform sizes as well as transform skip.

\begin{figure}
\centering
\subfigure[][]{
\includegraphics[width=0.46\linewidth]{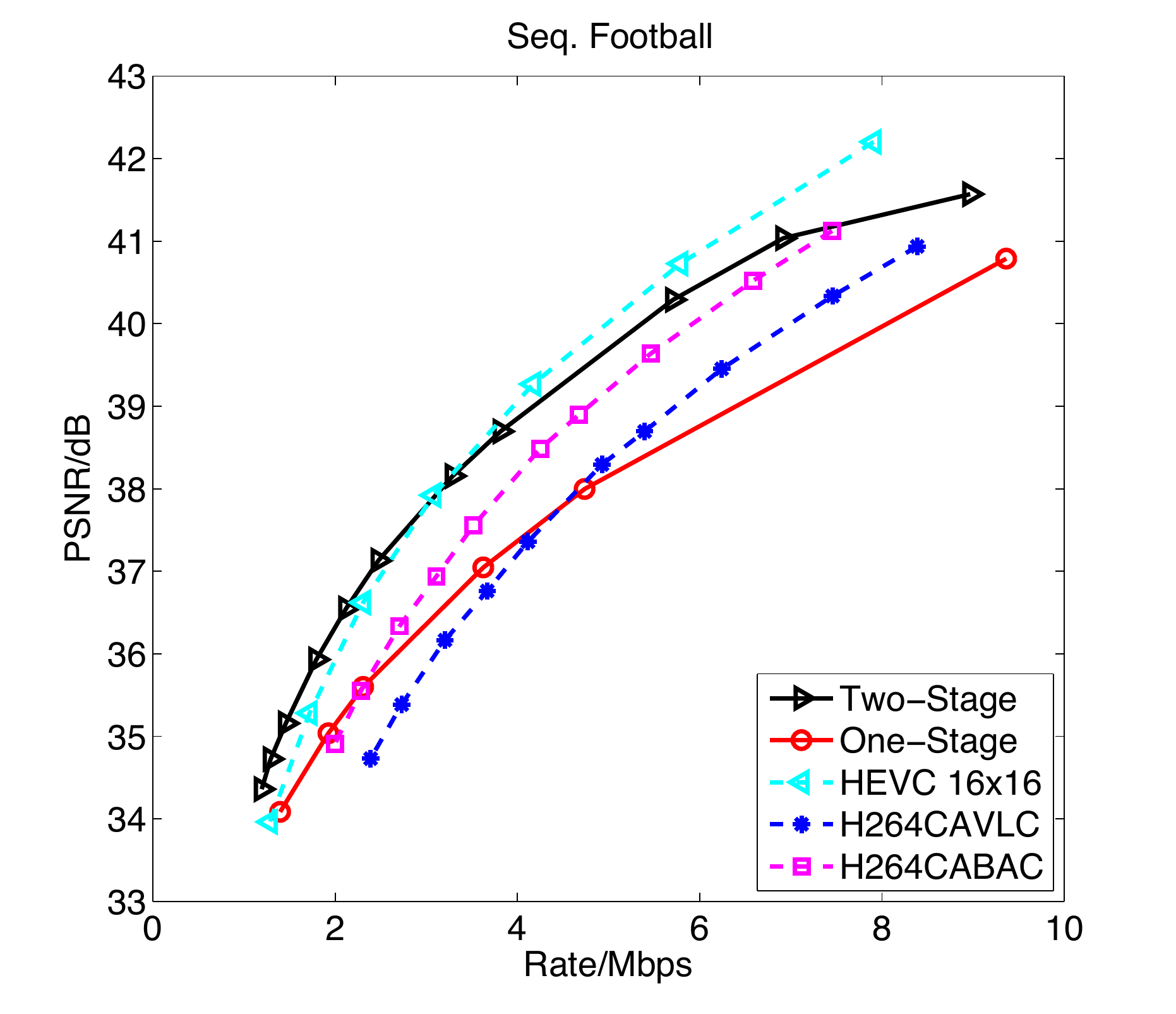}}
\subfigure[][]{
\includegraphics[width=0.46\linewidth]{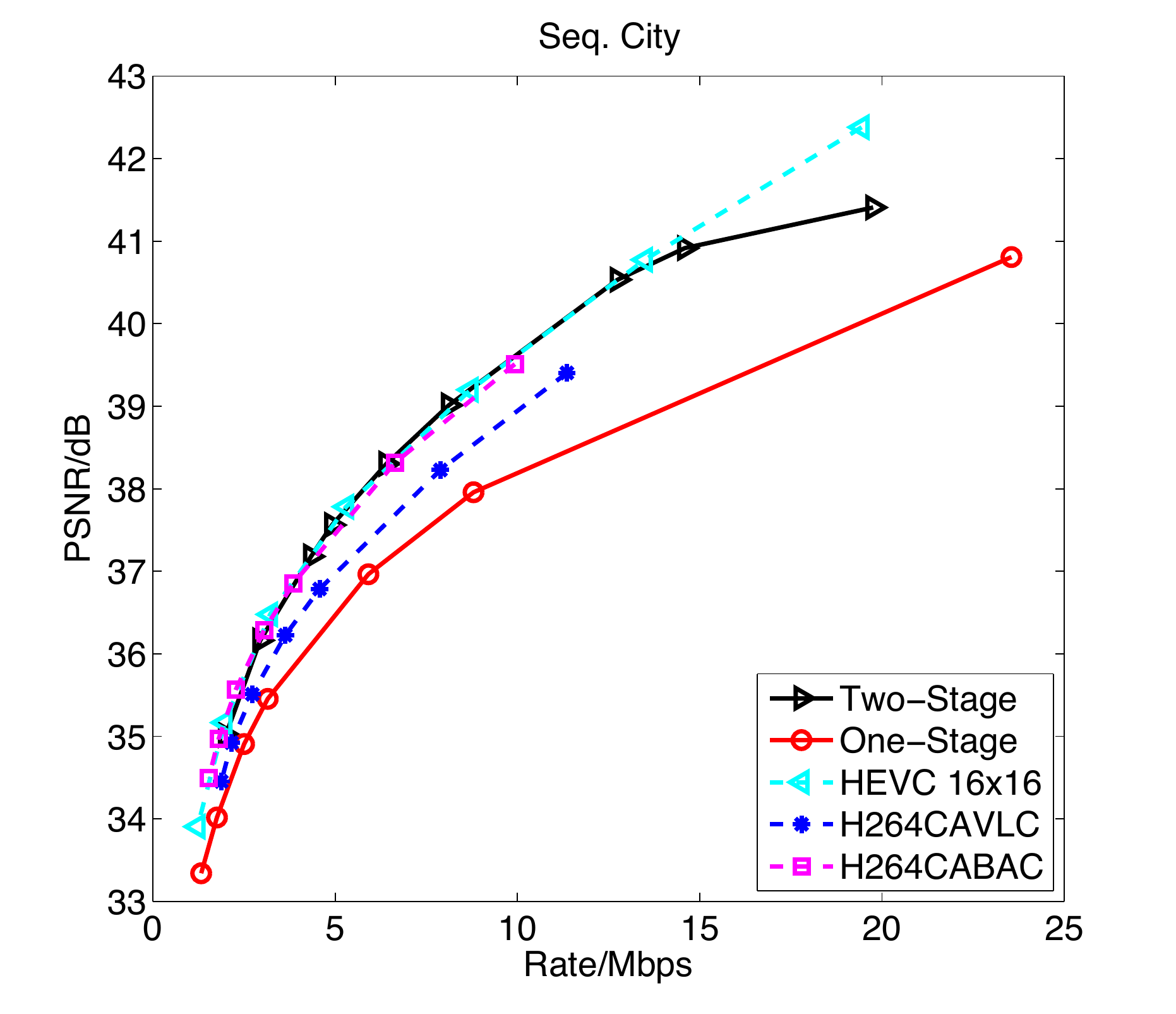}}
\caption{Ten P-frames Rate-distortion curves for proposed and comparison methods. a) Seq. \textit{football}, b) Seq. \textit{city}.}
\label{fig:rd10frm}
\end{figure}

Figure~\ref{fig:rd10frm} shows the rate-distortion curves for encoding ten P-frames, using IPPP structure. The rate and PSNR are averaged from all ten P-frames. 
Our proposed coder can still outperform both H.264 options, and is competitive with HEVC in the low to intermediate rate range (in fact the proposed coder is up to 0.5dB better for \textit{football} at low rate range). The significant gain that was achieved for coding the first P-frame at low rates vanished unfortunately. We suspect that this is because our coder has not implemented in-loop filters such as deblocking filter and sample adaptive offset filter. Those filters are known to mitigate the error propagation effect. Because our first stage dictionary consists of blocks from the previously decoded  frame, its representation power reduces when the previous frame has large quantization error. Effective in-loop filtering is likely to suppress this efficiency loss. Besides, our proposed coder only uses inter-prediction candidates in the self-adaptive first-stage dictionary. We speculate the absence of intra-modes may also contribute to the loss of efficiency in coding subsequent frames.
\section{Conclusion and Future work}\label{sec:con}
In this work we propose a two-stage video encoder framework, which uses a self-adaptive redundant dictionary in the first stage and uses DCT bases (after orthonormalization with the first stage chosen atoms) in the second stage. We use a residual norm reduction ratio threshold to switch from the first to the second stage. We further propose a complete context adaptive binary entropy coding method to efficiently code the location, order and the coefficient values of the chosen atoms of the first stage.  The location and values of non-zero coefficients in the second stage are coded following the same transform coding method of HEVC for coding the prediction residual, but using a fixed transform size of $16\times 16$. The two-stage coder achieved around 1dB PSNR improvement over our previous one-stage approach in~\cite{xue2014video}. Its performance is generally competitive with HEVC, with slightly better performance at low rates, and worse performance at very high rates. Note that our present coder uses a fixed block size of $16\times 16$, has only inter-prediction candidates in the self-adaptive dictionary, and has not implemented in-loop filtering.  The fact that it is competitive with HEVC is quite promising, considering that HEVC allows variable block size both for prediction and for coding of prediction residual, adpative switching between inter- and intra-prediction, and incorporates sophisticated in-loop filtering.  

There are many aspects of the present coder that can be improved. Adaptive switching from the first stage to the second stage on a block-basis is likely to yield significant gain. Incorporation of in-loop filtering and intra-prediction may also bring significant improvements.
%
\bibliographystyle{IEEEbib}
{
\bibliography{TwoStageICIP15}}

\begin{thebibliography}{10}

\bibitem{xue2014video}
Yuanyi Xue and Yao Wang,
\newblock ``Video coding using a self-adaptive redundant dictionary consisting
  of spatial and temporal prediction candidates,''
\newblock in {\em Multimedia and Expo (ICME), 2014 IEEE International
  Conference on}. IEEE, 2014, pp. 1--6.

\bibitem{HEVCreview}
Gary~J. Sullivan, Jens-Rainer Ohm, Woo-Jin Han, and Thomas Wiegand,
\newblock ``Overview of the high efficiency video coding (hevc) standard,''
\newblock {\em IEEE Trans. on Circuits and Systems for Video Technology}, vol.
  22, pp. 1649--1668, 2012.

\bibitem{KSVD_ImgComp}
Karl Skretting and Kjersti Engan,
\newblock ``Image compression using learned dictionaries by rls-dla and
  compared with k-svd,''
\newblock {\em IEEE International Conference on Acoustics, Speech, and Signal
  Processing}, 2011.

\bibitem{Guillemot_JSTSP_2011}
Joaquin Zepeda, Christine Guillemot, and Ewa Kijak,
\newblock ``Image compression using sparse representations and the
  iteration-tuned and aligned dictionary,''
\newblock {\em IEEE Journal of Selected Topics in Signal Processing}, vol. 5,
  no. 5, pp. 1061--1073, 2011.

\bibitem{Zakhor_Dictionary4vid_CSVT2004}
Philippe Schmid-Saugeon and Avideh Zakhor,
\newblock ``Dictionary design for matching pursuit and application to
  motion-compensated video coding,''
\newblock {\em IEEE Trans. on Circuits and Systems for Video Technology}, vol.
  14, no. 6, pp. 880--886, 2004.

\bibitem{MERL_ISCAS2011}
Je-Won Kang, C-CJ Kuo, Robert Cohen, and Anthony Vetro,
\newblock ``Efficient dictionary based video coding with reduced side
  information,''
\newblock in {\em Circuits and Systems (ISCAS), 2011 IEEE International
  Symposium on}. IEEE, 2011, pp. 109--112.

\bibitem{SparseDCT_TIP2013}
Je-Won Kang, Moncef Gabbouj, and C.-C.~Jay Kuo,
\newblock ``Sparse/dct (s/dct) two-layered representation of prediction
  residuals for video coding,''
\newblock {\em IEEE Trans. on Image Processing}, vol. 22, no. 7, pp.
  2711--2722, July 2013.

\bibitem{TsinghuaOnlineDictionary}
Yipeng Sun, Mai Xu, Xiaoming Tao, and Jianhua Lu,
\newblock ``Online dictionary learning based intra-frame video coding via
  sparse representation,''
\newblock in {\em Wireless Personal Multimedia Communications (WPMC), 2012 15th
  International Symposium on}, 2012.

\bibitem{xue2015revomp}
Yuanyi Xue and Yao Wang,
\newblock ``Eomp: Finding better representation by recursively orthonormalizing
  the remaining atoms,''
\newblock in {\em International Conference on Sampling Theory and
  Applications}. 2015, submitted.

\bibitem{OMP07}
Joel~A. Tropp and Anna~C. Gilbert,
\newblock ``Signal recovery from random measurements via orthogonal matching
  pursuit,''
\newblock {\em IEEE Trans. on Information Theory}, vol. 53, no. 12, pp.
  4655--4666, December 2007.

\bibitem{CABAC}
D.~Marpe, H.~Schwarz, and T.~Wiegand,
\newblock ``Context-based adaptive binary arithmetic coding in the h.264/avc
  video compression standard,''
\newblock {\em IEEE Trans. on Circuits and Systems for Video Technology}, vol.
  13, no. 7, pp. 620--636, July 2003.

\end{thebibliography}
\end{document}